----------------------------------------------------------------------------

\documentstyle[12pt]{article}
\begin{document}

\title{A new dynamical group approach \\to the Modified P\"{o}schl-Teller potential}

\author{Shi-Hai Dong \thanks{E-mail address: dongsh@nuclecu. unam. mx, lemus@servidor. unam. mx} and R. Lemus \\
{\small
Instituto de Ciencias Nucleares, UNAM, }\\
{\footnotesize  Apdo. Postal. 70-543, Circuito Exterior,
C. U. ,  04510 Mexico, D. F. , Mexico}}

\date{}

\maketitle

\vspace{4mm}

\begin{abstract}

The properties of the wave function for the
Modified P\"{o}schl-Teller (MPT) potential are outlined. The ladder operators are constructed
directly from the wave functions without introducing any auxiliary parameter.
It is shown that these operators are associated to the $su(2)$ algebra.
Analytical expressions for the
functions $\sinh(\alpha x)$ and $\frac{\cosh(\alpha x)}{\alpha} \frac{d}{dx}$ are
evaluated from ladder operators. The expansions of the coordinate $x$ and momentum $\hat p$ in terms of the $su(2)$ generators are presented. This analysis allows to establish an exact quantum-mechanical connection between  the $su(2)$ vibron model and the traditional descriptions of molecular vibrations.

\end{abstract}

\vskip 2cm

\noindent

PACS number(s): 03. 65. Ca, 03. 65. Fd








\newpage

\begin{center}

{\large{\bf 1. Introduction}}

\end{center}

During the past several decades, algebraic method has been applied to a wide variety of fields in
both physics and chemistry. Systems displaying a dynamical symmetry can be solved with
 algebraic techniques [1-4]. Particularly, the Morse [5] and P\"{o}schl-Teller (PT) potentials
[6, 7] represent two of the most studied anharmonic systems where these techniques have been used. Both of them are closely related with
the $SO(2, 1) $ [8-15] and $SU(2) $ groups [3, 4, 16-18]. The latter has been used to describe
the vibrational excitations of molecular systems, while the former is associated to
the potential group approach. The relation between the $SU(2)$ group and the Morse and PT
systems can be directly established by means of a coordinate
transformation applied to the radial equation of a 2D harmonic oscillator. The disadvantage of this approach need introduce an auxiliary variable to define the action of the group generators. It is possible, however, to obtain the ladder operators only in terms of the physical variable with the factorization method [16]. In this work, we shall present   an explicit
connection of the Modified P\"{o}schl-Teller (MPT) potential with the $su(2)$ algebra.

In the field of molecular physics, the PT potential was applied in the framework of the $su(2)$ vibron model, where it was associated
to the vibrational excitations of the molecular bending modes [19-21].
The basic idea behind the $su(2)$ vibron model consists in expanding the Hamiltonian in terms of different sets of $su(2)$ generators, whose number  corresponds to the internal vibrational degrees of freedom of the molecule. Generally, the stretching degrees of freedom are associated to a Morse potential but the bending coordinates can be identified with either the Morse or PT potentials. This model was empirically applied without establishing its connection  with the traditional approach in the configuration space. Consequently, in such a scheme the relation between the spectroscopic parameters and the molecular structure constants was lacking. Concerning the degrees of freedom associated to the Morse potential , this problem has been recently solved by establishing the relation between the matrix elements of two interacting Morse oscillators and the corresponding $SU(2)$ matrix elements. The $su(2)$ vibron model was shown to correspond to an approximation where only the dominant $\Delta v=\pm 1$ interaction between Morse oscillators is taken into account [17]. Moreover, a procedure to extend the algebraic approach in order to systematically approach the Morse oscillators results was proposed [18]. This procedure is based on the expansion of the coordinate $x$ and momentum $\hat p$ in terms of the $su(2)$ generators; an expansion that was obtained by comparison of the  matrix elements in both spaces. In this work we shall obtain the corresponding expansion of $x$ and $\hat p$ for the  MPT potential by inverting the expressions for the ladder operators. This analysis provides us for an approach to approximate the matrix elements of interacting PT oscillators in the framework of the $su(2)$ model.

This paper is organized as follows. The  properties of the MPT potential are outlined in Section 2.  In Section 3, the associated ladder operators are constructed directly from the wave functions with the factorization method [22-24] and their connection with the $su(2)$ algebra is established . In addition, the matrix elements for certain
relevant functions of the coordinate $x$ and momentum $\hat p$ are calculated from the  ladder operators. In Section 4 the expansions of the coordinate $x$ and momentum $\hat p$  in terms of the $su(2)$ generators are presented , including an analysis of two coupled PT oscillators system. Some concluding remarks will be given in Section 5.

\vskip 1cm

\begin{center}

{\large {\bf 2. The properties of the MPT potential}}

\end{center}

We start by presenting the solutions of the MPT problem [25]. The MPT potential as described in Ref. [26] can be written as

$$V(x)=-\frac{D}{\cosh^2(\alpha x)}, \eqno(1)$$
where $D$ is the depth of the well and $\alpha$ is related with the range of the potential, while $x$ gives the relative distance from the equilibrium position. The Schr\"{o}dinger equation associated to  this potential is given by
$$\frac{d^2\Psi_{n}^{q}(x)}{dx^2}+\frac{2\mu}{\hbar ^2}\left(E+\frac{D}{\cosh^2(\alpha x)}\right)\Psi_{n}^{q}(x)=0, \eqno(2)$$
where $\mu$ is the reduced mass of the molecule and $q$ is related with the depth of the potential. The solutions of Eq. (2) can be written as
$$\Psi_{n}^{q}(u)=N_{n}^{q}(1-u^2)^{\frac{\epsilon}{2}}C_{n}^{q+\frac{1}{2}-n}(u),~~
E_n=-\frac{\alpha ^2\hbar ^2}{2\mu}(q-n)^2,n=0,1,2,... \eqno(3)$$
where
$$\epsilon=\sqrt{\frac{-2\mu E}{\alpha ^2 \hbar ^2}}, q(q+1)=\frac{2\mu D}{\alpha ^2 \hbar ^2}, q=\frac{1}{2}(-1+2k), k=\sqrt{\frac{1}{4}+\frac{2\mu D}{\alpha ^2 \hbar ^2}}, \nu=2 k=2 q+1, \eqno(4)$$
where
$$u=\tanh(\alpha x),\epsilon=q-n>0, N_{n}^{q}=\sqrt{\frac{\alpha n!(q-n-\frac{1}{2})!(2q-2n)!}{\pi^{\frac{1}{2}}(q-n-1)!(2q-n)!}} \eqno(5)$$.
Here $\nu$ has been introduced because of its relevance  for the identification of the ladder operators with the $su(2)$ algebra, as will be shown in the next section.
On the other hand, the number of  bound states is determined by the dissociation limit  $\epsilon=q-n=0$.
We should note that for $q$ integer the state associated with null energy is not normalizable. In this case the last  bounded state corresponds to $q-n=1$. We thus have that $n_{max}=q-1=(\nu-3)/2$.

\vskip 1cm
\begin{center}

{\large {\bf 3. The construction of the ladder operators}}
\end{center}

In this section we address the problem of finding ladder operators with the factorization method , namely, we intend to find the differential operators $\hat {\cal P}$ with the following property

$$\hat {\cal P}_\pm \Psi_{n}^{q}(u) = p_\pm \Psi_{n\pm 1}^{q}(u). \eqno(6)$$
Specifically, we look for operators with the following form

$$\hat {\cal P}_\pm = A_\pm (u) {d \over du} + B_\pm (u), \eqno(7)$$
where we remark  that these operators only depend on the physical variable $u$.

The operators (14) can be found by obtaining the action of the differential operator $ d \over du $ on the MPT wave functions. Therefore the following formula  (see Eq. 8. 935 in [28])

$$ \frac{dC_{n}^{\lambda}(t)}{dt}=2\lambda C_{n-1}^{\lambda+1}(t), \eqno(8)$$
together with Eq. (3), allows to obtain

$$\frac{d\Psi_{n}^{q}(u)}{du}=-\frac{u(q-n)}{1-u^2}\Psi_{n}^{q}(u)+\frac{2q-2n+1}{\sqrt{1-u^2}}
\frac{N_{n}^{q}}{N_{n-1}^{q}}\Psi_{n-1}^{q}(u), \eqno(9)$$
and introducing the explicit form of the normalization constant, equation (9) becomes
$$\sqrt{1-u^2}\left(\frac{d}{du}+\frac{u(q-n)}{1-u^2}\right)\sqrt{\frac{q-n+1}{q-n}}
\Psi_{n}^{q}(u)=\sqrt{n(2q-n+1)}\Psi_{n-1}^{q}(u), \eqno(10)$$
from which we can define the annihilation operator $\hat P_{-}$ as

$$\hat P_-=\sqrt{1-u^2}\left(\frac{d}{du}+\frac{u(q-n)}{1-u^2}\right)\sqrt{\frac{q-n+1}{q-n}}, \eqno(11)$$
or in terms of $\nu$ defined in (5)
$$\hat P_-=\sqrt{1-u^2}\left(\frac{d}{du}+\frac{u}{1-u^2}~~ \epsilon \right)\sqrt{\frac{\epsilon+1}{\epsilon}}, \eqno(12)$$
where in order to simplify the notation we have taken into account  that $2\epsilon=\nu-2n-1=2q-2n$. The
action of the operator (12)  on the wave functions is then given by

$$\hat P_{-}\Psi_{n}^{\nu}(u)=p_-\Psi_{n-1}^{\nu}(u), \eqno(13)$$
where

$$p_-=\sqrt{n(\nu-n)}. \eqno(14)$$
As we can see, this operator annihilates the ground state $\Psi_{0}^{\nu}(u)$, as expected
from a lowering operator.

We now proceed to find the corresponding creation operator $\hat P_{+} $. To this end, we consider the  formula (see (10. 43) of [29])
$$2(\lambda-1)(2\lambda-1)xC_n^{\lambda}(x)=4\lambda(\lambda-1)(1-x^2)C_{n-1}^{\lambda+1}(x)+(2\lambda+n-1)(n+1)C_{n+1}^{\lambda-1}(x). \eqno(15)$$
This recurrence relation can be used together with Eq. (8) to obtain
$$\frac{d\Psi_{n}^{q}(u)}{du}=\frac{u(q-n)}{1-u^2}\Psi_{n}^{q}(u)
-\frac{(n+1)(2q-n)}{\sqrt{1-u^2}(2q-2n-1)}\frac{N_n^q}{N_{n+1}^q}\Psi_{n+1}^{q}(u). \eqno(16)$$
By using the explicit form of the normalization constant (5), equation (16) becomes

$$-\sqrt{1-u^2}\left(\frac{d}{du}-\frac{u(q-n)}{1-u^2}\right)
\sqrt{\frac{(q-n-1)}{(q-n)}}\Psi_{n}^{q}(u)=\sqrt{(n+1)(2q-n)}
\Psi_{n+1}^{q}(u). \eqno(17)$$
Likewise, in terms of the variable $\nu$, we can thus define the  creation operator $\hat P_+$ as

$$\hat P_{+}=\sqrt{1-u^2}\left(-\frac{d}{du}+\frac{u}{1-u^2}~~\epsilon\right)\sqrt{\frac{\epsilon-1}{\epsilon}}, \eqno(18)$$
with the following effect on the wave functions

$$\hat P_+\Psi_{n}^{\nu}(u)=p_+\Psi_{n+1}^{\nu}(u), \eqno(19)$$
where

$$p_{+}=\sqrt{(n+1)(\nu-n-1)}. \eqno(20)$$
From Eqs. (8) and (17) we see the importance of normalization factors. Since $\hat P_+$ is a raising operator it is expected to annihilate the last bounded state. Indeed, for such state $\epsilon=1$ and the square root in (18) makes the operator vanish.

We now establish the algebra associated  with the operators $\hat P_{\pm} $. Based on the Eqs. (12) and (19), we calculate the commutator $[\hat P_{-}, \hat P_{+}]$:

$$[\hat P_{+}, \hat P_{-}] \Psi_{n}^{\nu}(u)=2 p_0 \ \Psi_{n}^{\nu}(u) \eqno(21)$$
where we have introduced the eigenvalue

$$p_{0}=-\left(\frac{\nu-1}{2}-n \right). \eqno(22)$$
We can thus define the operator

$$\hat P_{0}=\hat n -\frac{\nu-1}{2}. \eqno(23)$$

The operators $\hat P_{\pm, 0}$ satisfy the commutation relations

$$[\hat P_{+}, \hat P_{-}]=2\hat P_0, ~~[\hat P_{0}, \hat P_-]=-\hat P_{-}, ~~[\hat P_{0}, \hat P_+]=\hat P_{+}, \eqno(24)$$
which correspond to the $su(2)$ algebra. This result is consistent with the description of
a finite discrete spectrum, in accordance with previous algebraic descriptions of the
bounded states of the P\"{o}schl-Teller potential [4, 8].
The number of bosons $N$ is related with $\nu$ by $N=\nu-1$, as we deduce from the Casimir operator

$$\hat C \ \Psi_{n}^{\nu}(u)=\left[ \hat P^2_0+{1\over 2} (\hat P_{+} \hat P_{-} + \hat P_{-} \hat P_{+} )\right] \Psi_{n}^{\nu}(u) =j(j+1)\Psi_{n}^{\nu}(u), \eqno(25)$$
where $j$, the label of the irreducible representations of su(2), is given by

$$j={\nu-1\over 2}={N\over 2}. \eqno(26)$$
From the commutation relations (24), we know that $\hat P_0$ is the projection of the angular momentum $m$, and consequently

$$n-{\nu-1\over2}=m. \eqno(27)$$
The ground state thus corresponds to $m=-j$, while the maximum number of quanta $n_{max}=
{\nu-3\over2}$ and consequently $m_{max}|_{n_{max}}=-1$, in accordance with the constraint
condition $\epsilon=q-n=1$ for the last bounded state. The MPT wave functions are thus associated  to one branch (in this case to $m\leq -1$) of the su(2) representations, as expected.  Finally we should notice that in terms of the $su(2)$ algebra, the Hamiltonian acquires the simple form

$$\hat H=- \frac{\hbar \omega}{\nu}\hat P_0^2, \eqno(28)$$
where
$$
\omega=\frac{\hbar \beta^2 \nu}{2 \mu}.$$

while for the wave functions

$$\Psi_{n}^{\nu}(u)={\cal N}_{n}^{\nu} {\hat P_{+}}^n \Psi_{0}^{\nu}(u), \eqno(29)$$
where the normalization constant is obtained through the commutation relations (24), and turns out to be

$${\cal N}_{n}^{\nu} =\sqrt{{(\nu-n-1)!\over {n! (\nu-1)!}} }. \eqno(30)$$

For other calculations one can obtain the following expressions in terms of the raising and lowering  operators $\hat P_{\pm}$

$$\frac{u}{\sqrt{1-u^2}}=\frac{1}{2}\left(\hat P_- \sqrt{\frac{1}{\epsilon(\epsilon+1)}}+
\hat P_+ \sqrt{\frac{1}{\epsilon(\epsilon-1)}}\right), \eqno(31)$$

$$\sqrt{1-u^2}\frac{d}{du}=\frac{1}{2}\left(\hat P_-\sqrt{\frac{\epsilon}{\epsilon+1}}-\hat P_+
\sqrt{\frac {\epsilon}{\epsilon-1}}\right), \eqno(32)$$
where it has to be understood that for the last bounded state ($\epsilon=1$) the raising operator vanishes.
On the other hand, we remark that the variable $\epsilon$ is to be considered as an $n$ dependent operator. This is the price to pay for having ladder operators in terms of the physical variable $u$, which  contrasts with the 2D description, where no number operators are involved but an auxiliary variable must be included [13].
Using Eqs. (13) and (19) and considering the constraint condition $2\epsilon=\nu-2n-1$, we can thus calculate the matrix elements of these functions as

$$\begin{array}{ll}<\Psi^{\nu}_{n^{\prime}}(u)|\frac{u}{\sqrt{1-u^2}}|\Psi^{\nu}_{n}(u)>
&=~<\Psi^{\nu}_{n^\prime}(x)|\sinh(\alpha x)|\Psi^{\nu}_{n}(x)>\\
&=~\displaystyle{\sqrt{\frac{n(\nu-n)}{(\nu-2n-1)(\nu-2n+1)}}}\delta_{n^{\prime}, n-1}\\
&+~\displaystyle{\sqrt{\frac{(n+1)(\nu-n-1)}{(\nu-2n-1)(\nu-2n-3)}}}\delta_{n^{\prime}, n+1}

\end{array}, \eqno(33)$$

$$\begin{array}{ll}<\Psi^{\nu}_{n^\prime}(u)|\sqrt{1-u^2}\frac{d}{du}|\Psi^{\nu}_{n}(u)>
&=~<\Psi^{\nu}_{n^\prime}(x)|\frac{\cosh(\alpha x)}{\alpha} \frac{d}{dx}|\Psi^{\nu}_{n}>(x)\\
&=~\displaystyle{\frac{1}{2}\sqrt{\frac{n(\nu-n)(\nu-2n-1)}{(\nu-2n+1)}}}\delta_{n^{\prime}, n-1}\\
&-~\displaystyle{\frac{1}{2}\sqrt{\frac{(n+1)(\nu-n-1)(\nu-2n-1)}{(\nu-2n-3)}}}\delta_{n^{\prime}, n+1}
\end{array}. \eqno(34)$$

\vskip 1cm

\begin{center}

{\large {\bf 5. Connection with the $su(2)$ vibron model}}

\end{center}

In the previous sections it is showed that the $su(2)$ algebra can be identified as the dynamical group for the MPT potential. Introducing the following renormalized expressions

$$\hat b ^\dagger={\hat P_+ \over \sqrt{\nu} }\ , ~~ \hat b={\hat P_- \over \sqrt{\nu} }\ , ~~ \hat b_0={\hat P_0 \over \nu}. \eqno(35)$$
we can obtain the simplified MPT wave functions

$$\Psi_{n}^{\nu}(u)=\sqrt{{\nu^n (\nu-n-1)!\over {n! (\nu-1)!}} }
 \ ( \hat b ^\dagger)^n \ \Psi_0^\nu(u), \eqno(36)$$
which implies that he wave functions can be expressed in terms of a repeatedly action of the creation operator $\hat b ^\dagger$ on the ground state. On the other hand, it was well known the connection of the PT potential with the $SU(2)$ group through a 2D harmonic oscillator. This connection  allows to associate the PT potential to certain  bending degrees of freedom (e.g. the out of plane local modes) in the framework of the $su(2)$ vibron model. This model consists in expanding the Hamiltonian in terms of $su(2)$ generators associated to  local oscillators (internal coordinates). It is not obvious, however, the relation of such generators with the creation and annihilation operators obtained in Section 3. The aim of this section is to establish this connection.

Let us consider  two interacting  oscillators
in the framework of the $su(2)$ vibron model.  According to this
approach the Hamiltonian has the form [18]
$$
H_{su(2)} = {\hbar \over 2} \omega_0 \sum^2_{ i = 1} (\hat b ^\dagger_i
\hat b_i + \hat b_i \hat b ^\dagger_i) + \lambda \hbar \omega_0 ( \hat b ^\dagger_1 \hat b_2 +
\hat  b_1 \hat b ^\dagger _2 ) ~~ . \eqno(37)
$$
The first term corresponds to the independent oscillators identified with either the Morse or PT potentials. In this study, we shall consider PT potential. The parameter $\omega_0$ and the
boson number $N$ are  related to the PT spectroscopic parameters through [18]

$$
\omega_e = \hbar \tilde \omega = \hbar\omega_0 { ( N+1) \over N } ~~
, \qquad x_e \omega_e = {\hbar\omega_0 \over  N} ~~ , \eqno(38)
$$
where

$$
\tilde \omega=\frac{\omega_e} {\hbar}=\frac{\alpha^2 \hbar} {\mu} \sqrt{\frac{1} {4}+\frac{2 \mu D} {\alpha^2 \hbar^2}}; ~~~~\qquad x_e \omega_e =\frac{\alpha^2 \hbar^2} {2 \mu}. \eqno(39)
$$
The diagonal terms generate PT (Morse)-like spectra, while the interaction term
$$
H^{su(2)}_{int.} = \lambda \hbar \omega_0 \, (\hat b ^\dagger_1 \hat b_2 + \hat  b_1
\hat b ^\dagger_2 ) ~~ , \eqno(40)
$$
gives rise to  the   matrix elements
$$
 < N , n_1 +1,  N, n_2 -1 | H^{su(2)}_{int.} | N n_1,
N n_2>  = \lambda \hbar \omega_0 \sqrt{n_2 (n_1 +1)} \, \sqrt{ \left[1 - {(n_2
-1) \over N} \right] \left[ 1 - {n_1 \over N}\right] } ~~ , \eqno(41)
$$
where $n_i$ stands for the number of quanta of the $i$-th oscillator. In the $su(2)$ vibron model the local functions are denoted by $|N, n>$. We note that  corrections of order $1/N$ appear.   When $N\to \infty$ the
matrix elements (41) reduce to
$$
\lim_{N \rightarrow \infty} < n_1 + 1, n_2 - 1 | \hat H^{su(2)}_{int.} | n_1, n_2> = \lambda \hbar \omega_0
\sqrt{ n_2 (n_1 + 1) } ~~ , \eqno(42)
$$
as expected, since in this limit all the harmonic results are recovered [18]. In order to interpret the correction appearing in (42), we shall analyze two PT oscillators from the point of view of the dynamical group identified in Section 4.

The Hamiltonian of two identical MPT oscillators in configuration
space  is given by
$$
H = {1\over 2\mu} \, \sum^2 _{ i = 1 } \hat {p}^2_i - D \sum^2_{ i = 1 } \frac{1}{\cosh^2(\alpha x)}+ \lambda \left( {\hat p_1 \hat p_2 \over \mu} + \mu
\tilde\omega^2 r_1
r_2\right) ~~ .  \eqno(43)
$$
The first two terms correspond to two PT independent oscillators, while the third is a particular interaction chosen  to correspond  in the limit $N \rightarrow \infty $ to the interaction$$
\hat {H}^{h.o.}_{int.}  = \hbar \lambda \omega \
(\hat a^\dagger_1 \hat a_2 + \hat a_1 \hat a^\dagger_2) ~~   \eqno(44)
$$
for two harmonic oscillators. In configuration space Eq.(44) takes the equivalent
form
$$
H^{h.o.}_{int.} =   \lambda \left( {\hat p_1 \hat p_2 \over \mu} + \mu \omega^2
x_1 x_2\right)~~, \eqno(45)
$$
with $\omega$ given by
$$
\omega=\sqrt{f_{rr} g_{rr}}.\eqno(46)
$$
Note that the matrix elements of the interaction (44)  coincide with the matrix elements in (42).

We now address the problem of computing the non diagonal contributions
to the interaction term
$$
H^{MPT}_{int.} = \lambda \left( {\hat p _1 \hat p_2 \over \mu} + \mu \tilde\omega^2 x_1 x
_2 \right) ~~ . \eqno(47)
$$
To this end we introduce the bosonic operators
$$
 \hat  c^\dagger_i  = \sqrt{{\mu\tilde\omega \over 2\hbar}} x_i - i
\sqrt{{ 1 \over 2\hbar \tilde \omega \mu}} \hat p_i ~~ , \eqno(48a)
$$

$$
\hat c_i = \sqrt{{ \mu \tilde\omega \over 2 \hbar}} x_i + i \sqrt{{ 1 \over 2
\hbar \tilde\omega \mu}} \hat p_i  ~~ .\eqno(48b)
$$
In terms of these operators the
interaction (47) acquires the simple form
$$
H^{MPT}_{int.} = \hbar \tilde\omega \lambda (\hat  c^\dagger_1 \hat c_2 + \hat c_1
\hat  c^\dagger _2)
~~ . \eqno(49)
$$
In order to establish the relation between this term and the interaction associated to the $su(2)$ vibron model given by (40), we proceed to express (49) in terms of the $su(2)$ ladder operators obtained in Section 3. This task can be achieved by extracting the coordinate $x$ and momentum  $\hat p$ from (31) and (32), which implicitly are given by

$$\sinh{\alpha x}=\frac{1}{2}\left(\hat b \sqrt{\frac{\nu}{\epsilon(\epsilon+1)}}+
\hat b ^\dagger \sqrt{\frac{\nu}{\epsilon(\epsilon-1)}}\right), \eqno(50)$$

$$\frac{\cosh(\alpha x)}{\alpha} \frac{d}{dx}=\frac{1}{2}\left(\hat b\sqrt{\frac{\nu \epsilon}{\epsilon+1}}-\hat b ^\dagger
\sqrt{\frac {\nu \epsilon}{\epsilon-1}}\right). \eqno(51)
$$
If we now take into account the Taylor series
$$
\hbox{arcsinh} {(x)}=x-\frac{x^3}{6}+ \frac{3 x^5} {40}+ ..., \eqno(52)
$$
$$
\hbox{sech} {(x)}=1-\frac{x^2}{2}+ \frac{5 x^4} {24}+ ..., \eqno(53)
$$
keeping up to the second terms, we obtain from (50) an (51) the following expansions
\begin{eqnarray}
\alpha {x} & = & \frac {1} {2} (\hat b ^\dagger f_n + \hat b g_n)-\frac{1} {48} (
 \hat b ^\dagger f_n \hat b ^\dagger f_n \hat b ^\dagger f_n
+\hat b ^\dagger f_n \hat b ^\dagger f_n \hat b g_n
+\hat b ^\dagger f_n \hat b g_n \hat b ^\dagger f_n
+\hat b ^\dagger f_n \hat b g_n \hat b g_n \nonumber \\
& + & \hat b g_n \hat b ^\dagger f_n \hat b ^\dagger f_n +\hat b g_n \hat b ^\dagger f_n \hat b g_n
+\hat b g_n \hat b g_n \hat b ^\dagger f_n +\hat b g_n \hat b g_n \hat b g_n) +...\nonumber
\end{eqnarray}
\vskip -20pt
$$\eqno{(54)}$$
\begin{eqnarray}
\hat p & = & \frac {i \hbar \alpha} {2} (\hat b ^\dagger h_n - \hat b q_n)+\frac{i \hbar \alpha} {16} (
 \hat b ^\dagger f_n \hat b ^\dagger f_n \hat b ^\dagger h_n
-\hat b ^\dagger f_n \hat b ^\dagger f_n \hat b q_n
+\hat b ^\dagger f_n \hat b g_n \hat b ^\dagger h_n
-\hat b ^\dagger f_n \hat b g_n \hat b q_n \nonumber \\
& + & \hat b g_n \hat b ^\dagger f_n \hat b ^\dagger h_n -\hat b g_n \hat b ^\dagger f_n \hat b q_n
+\hat b g_n \hat b g_n \hat b ^\dagger h_n +\hat b g_n \hat b g_n \hat b q_n )+..., \nonumber
\end{eqnarray}
\vskip -20pt
$$\eqno{(55)}$$
where we have introduced the notations

$$
f_n=\sqrt{\frac{\nu} {\epsilon (\epsilon-1)}}, ~~~~~g_n=\sqrt{\frac{\nu} {\epsilon (\epsilon+1)}} \eqno(56)
$$

$$
h_n=\sqrt{\frac{\nu \epsilon} { \epsilon-1}}, ~~~~~q_n=\sqrt{\frac{\nu \epsilon} { \epsilon+1}}, \eqno(57)
$$
as well as  $\hat p=-i \hbar \frac{d}{dx}$. We recall that the $n$ dependence if these variables are given through $2 \epsilon=\nu-2 n-1$. Note that because of the symmetry of the MPT potential
the expansions of $x$ and $\hat p$ involve only odd powers of the operators $\hat b ^\dagger (\hat b)$; $x$ and $\hat p$ are odd functions under inversion.
If we introduced the expressions (54) and (55) in the definitions (48), we would obtain an expansion of  $\hat  c^\dagger (\hat c)$ in terms of the $\hat b ^\dagger (\hat b)$ operators. As a first approximation, however, we shall consider contributions up to  linear terms in (54) and (55). If this is the case  the operators (48) take the approximate form
$$
 \hat  c^\dagger  \simeq \sqrt{{\mu\tilde\omega \over 2\hbar}} \left[ \frac {1} {2 \alpha} (\hat b ^\dagger f_n + \hat b g_n)\right] +
\sqrt{{ 1 \over 2\hbar \tilde \omega \mu}} \left[\frac { \hbar \alpha} {2 } (\hat b ^\dagger h_n- \hat b q_n )\right]~~ , \eqno(58a)
$$

$$
\hat  c \simeq \sqrt{{\mu\tilde\omega \over 2\hbar}} \left[ \frac {1} {2 \alpha} (\hat b ^\dagger f_n + \hat b g_n)\right] -
\sqrt{{ 1 \over 2\hbar \tilde \omega \mu}} \left[\frac { \hbar \alpha} {2} (\hat b ^\dagger h_n- \hat b q_n )\right]~~ , \eqno(58b)
$$
which can be rearranged in the form
$$
\hat  c^\dagger \simeq \hat b ^\dagger ~z_n + \hat b~ \zeta_n , \eqno(59a)
$$
$$
\hat c \simeq \hat b ^\dagger~ \zeta_n+ \hat b~z_n   , \eqno(59b)
$$
where we have introduced the functions
$$
z_n=\frac{1}{2}  \left( \sqrt{\frac{1}{ (1-(2 n+1)/\nu) (1-(2 n+3)/\nu)}    }+
\sqrt{\frac{1-(2 n+1)/\nu}{1-(2 n+3)/\nu}} \right), \eqno(60)
$$
$$
\zeta_n=\frac{1}{2}  \left( \sqrt{\frac{1}{ (1-(2 n+1)/\nu) (1-(2 n-1)/\nu)}    }-
\sqrt{\frac{1-(2 n+1)/\nu}{1-(2 n-1)/\nu}} \right).\eqno(61)
$$
It is clear that these functions have the harmonic limit
$$
\lim_{\nu \rightarrow \infty} z_n=1, ~~~~~~\lim_{\nu \rightarrow \infty} \zeta_n=0, \eqno(62)
$$
although for small number of quanta this limit represents a good  approximation. Only when $n$ takes its highest value this approximation fails. Hence we have the following limit$$
\lim_{ n \ll n_{max}} \hat  c^\dagger=\hat b ^\dagger, ~~~~~~\lim_{ n \ll n_{max}} \hat c=\hat b.\eqno(63)
$$
Let us now return to the two oscillators system. Considering the approximation (59), we obtain
$$
 (\hat c^\dagger_1 \hat c_2 + \hat c_1
\hat  c^\dagger _2) \simeq ( \hat b ^\dagger_1 \hat b_2 + \hat  b_1 \hat b ^\dagger _2 ) \hat A(n_1, n_2)+( \hat b ^\dagger_1 \hat b ^\dagger_2 + \hat  b_1 \hat b_2 ) \hat B(n_1, n_2)
~~ . \eqno(64)
$$
where
$$
\hat A(n_1, n_2)=z_{n_1} ~z_{n_2}+\zeta_{n_1}~\zeta_{n_2}, \eqno(65)
$$
$$
\hat B(n_1, n_2)=z_{n_1} ~\zeta_{n_2}+z_{n_2}~\zeta_{n_1}.\eqno(66)
$$
From the discussion for one oscillator, a reasonable approximation for small $n_1$ and $n_2$ is the following
$$
 \hat A(n_1, n_2) \simeq 1, ~~~~~~ \hat B(n_1, n_2) \simeq 0, \eqno(67)
$$
which implies $$
H^{MPT}_{int.} = \hbar \tilde\omega \lambda (\hat  c^\dagger_1 \hat c_2 + \hat c_1
\hat  c^\dagger _2) \simeq \hbar \tilde\omega \lambda ( \hat b ^\dagger_1 \hat b_2 + \hat  b_1 \hat b ^\dagger _2 )
~~ .\eqno(68)
$$
This expression corresponds essentially to the $su(2)$ vibron interaction (40).
We thus conclude that the $su(2)$ vibron model very nearly corresponds to taking the dominant $\Delta n=\pm 1$ interaction between coupled MPT oscillators, as long as we exclude the neighborhood of the dissociation energy.

This analysis suggests the following two steps to improve the the $su(2)$ vibron model. First we can consider the operators (59) instead of the single generators $\hat b ^\dagger$ and $\hat b$. The next step would include the next terms in the expansions (54) and (55). Both steps imply the breaking of the polyad $P=n_1+n_2$ in a specific way, specially for high number of quanta. This approach may thus be useful for the descriptions of vibrational excitations near the dissociation limit.

\vskip 1cm

\begin{center}

{\large {\bf 5. Concluding remarks}}

\end{center}

In this work we have outlined the properties of the MPT wave functions and  established the ladder operators with the factorization method. The realization obtained for the creation and annihilation operator, given in terms of the physical variable $u$, has been identified with the $su(2)$ algebra. This result is in accordance with  previous analysis in which the bounded region of the MPT potential is associated with an $su(2)$ algebra. The identification of the ladder
operators as generators of the  $SU(2)$ group allows to  express the MPT
wave functions in a simple closed form in terms of the action of $\hat b ^\dagger$ on the ground state. The matrix elements for the functions $\sinh(\alpha x)$ and $\frac{\cosh(\alpha x)}{\alpha} \frac{d}{dx}$ have been analytically obtained from the ladder operators $\hat P_{\pm, 0}$. This method represents a simple and elegant
approach to obtain these matrix elements.

We have also analyzed the theoretical relationship between the $su(2)$ vibron model and the coupled MPT system. It has been shown that the standard  $su(2)$ approach to molecular vibrational excitations (vibron model) is equivalent to considering MPT oscillators via the dominant $\Delta n= \pm 1$ selection rule. This approximation arises naturally from an expansion of the PT operators $\hat  c^\dagger (\hat c)$  in terms of the $su(2)$ generators $\hat b ^\dagger (\hat b)$, which in turn allows to establish an extended $su(2)$ model when the next terms in the expansion are taken into account. The extended $su(2)$ model may provide an approach to the vibrational description of the high energy region of the spectra, where polyad is breaking and  localization takes over the simple normal behavior at the low end of the spectrum.

\vspace{10mm}

\noindent
{\large {\bf Acknowledgments}}.

We are indebted to A.Frank for invaluable discussions and suggestions. This work is supported by CONACyT, Mexico, under project 32397-E.

\end{document}